\documentclass[%
 twocolumn,
 aps,
 showkeys,
 amsmath,amssymb,
]{revtex4-2}

\usepackage[a4paper,
            bindingoffset=0.0in,
            left=1in,
            right=1in,
            top=1in,
            bottom=1in,
            footskip=.25in]{geometry}
            
\usepackage{graphicx}% Include figure files
\usepackage{dcolumn}% Align table columns on decimal point
\usepackage{bm}% bold math
\usepackage{float}
\usepackage{mathtools}
\usepackage{xcolor}
\usepackage{soul}
\usepackage{ulem}
\usepackage{amsmath}
\usepackage{commath}
\usepackage{hyperref}
\usepackage{subcaption}

\DeclarePairedDelimiterX\braket[2]{\langle}{\rangle}{#1 \delimsize\vert #2}

\usepackage[ruled,vlined]{algorithm2e}
\usepackage{lipsum}
\makeatletter
\renewcommand\frontmatter@abstractwidth{\dimexpr\textwidth-1.5in\relax}
\makeatother

\begin{document}

\title{Application of Unsupervised Domain Adaptation for Structural MRI Analysis}

\author{Pranath Reddy}
\affiliation{University of Florida, Gainesville, FL 32611, USA}

\begin{abstract}
The primary goal of this work is to study the effectiveness of an unsupervised domain adaptation approach for various applications such as binary classification and anomaly detection in the context of Alzheimer’s disease (AD) detection for the OASIS datasets \cite{10.1162/jocn.2009.21407} \cite{10.1162/jocn.2007.19.9.1498}. We also explore image reconstruction and image synthesis for analyzing and generating 3D structural MRI data to establish performance benchmarks for anomaly detection. We successfully demonstrate that domain adaptation improves the performance of AD detection when implemented in both supervised and unsupervised settings. Additionally, the proposed methodology achieves state-of-the-art performance for binary classification on the OASIS-1 dataset.
\end{abstract}

\keywords{ML in Medicine, Deep Learning, Classification, Anomaly Detection, Domain Adaptation}

\maketitle

\section{Introduction}
Alzheimer's disease (AD), one of the most prevalent degenerative conditions, is a degenerative dementia that begins with minor memory loss and gradually escalates to a complete loss of mental and physical capacities. For the patient's health, a diagnosis should be made as soon as possible so that treatment and preventative measures can be commenced. A thorough and comprehensive medical evaluation involving an array of psychological and physical testing is necessary to diagnose AD. The ability and expertise of the clinician have a direct impact on the precision of the psychological and cognitive tests, and a magnetic resonance imaging (MRI) medical data analysis is necessary for a conclusive medical diagnosis of AD. Medical specialists are in responsibility of evaluating and interpreting medical data, however due to the data subjectivity and complexity, this procedure is particularly complex and constrained for a medical specialist. Consequently, there is a need for the development of quick and efficient diagnostic tools, and machine learning (ML) can be employed for this.

The ability of the computer to learn from experience without explicit programming has improved thanks to machine learning approaches. For classification, regression, clustering, and dimensionality reduction in a variety of applications like image processing, predictive analytics, and data mining, machine learning approaches are applied. In the field of medical imaging research, machine learning (ML) is frequently used and produces outstanding results in tasks like segmentation, regression, classification, etc. Medical image analysis tasks seldom have access to such expansive datasets, unlike some visual image datasets like ImageNet, which offer large-scale annotated examples for developing models for tasks like object detection. Using pre-trained models for some related domains using the transfer learning method is a practical and effective solution.

This study deals with a specific type of transfer learning approach called domain adaptation. Domain adaptation is a very effective learning technique that has been used for analyzing medical image data \cite{2021arXiv210209508G}. In this project, we will be focusing on studying 3D structural MRI brain scans in the context of Alzheimer’s disease classification. Real-world medical datasets are prone to being filled with unlabeled and ill-labeled samples, which makes working with them very challenging. The process of labeling medical images is typically expensive, time-consuming, and labor-intensive, necessitating the involvement of doctors, radiologists, and other experts. Where transfer learning strategies like domain adaptation may be useful is in this situation. While there has been a lot of work related to the development of novel classification algorithms for Alzheimer’s detection and classification, the study of interoperability and reusability of such models is often ignored, and this is very important since a model that performs well on one dataset might not replicate that performance when tested on a different dataset, and the samples themselves can be very different with a large domain gap between them, and domain adaptation could be useful in mitigating problems associated with domain shift.

\section{Background}

Here we discuss some of the prior work done in the domain of AD detection using machine learning. Detecting Alzheimer’s disease in its early stages can have a positive impact on the patient. To accomplish this, machine learning models have been at the forefront. Authors in \cite{Muhammed_Raees_2021} use the well-known ADNI (Alzheimer’s Disease Neuroimaging Initiative) dataset \cite{Jack2008-ia} to train, test, and compare SVM and different Deep Neural Networks (DNN), showcasing how the newer DNN models (VGG19) without lesser data preprocessing can achieve better accuracy of 90\% and more when compared to SVMs. The authors in \cite{Liu2022} employed 3D CNNs in conjunction with structural MRIs to accurately distinguish between those with moderate cognitive impairment, those with cognitive health, and those with mild dementia brought on by Alzheimer's disease. The authors of \cite{Odusami2021-np} trained ResNet18, an 18-layer CNN architecture, to predict various stages of cognitive impairment, and the model was able to achieve classification accuracy of near unity.

These texts mostly deal with using the entire image to train and predict. On the other hand, texture and shape features of the hippocampus region specifically are extracted, and a neural network acts as a multiclass classifier in \cite{Raut2017AML}. This approach is believed by the authors to yield better results. Using more than one model, as in ensemble learning (EL), has been shown to improve results and learning system performance. The authors of \cite{10.3389/fnins.2020.00259} combine CNN and EL to identify subjects with mild cognitive impairment (MCI) or AD. The authors in \cite{OROUSKHANI2022100066} employ a deep triplet network and a few-shot learning method, which surpass state-of-the-art models in terms of accuracy scores. This model addresses the issue of overfitting due to the limited image samples being used to train. After comparing many models on the OASIS dataset, the authors \cite{Battineni2021-rt} conclude that the gradient boosting technique may be a better classifier than others.

\section{Overview}

We will be studying AD detection in the context of both binary classification and anomaly detection. Binary classification is a supervised learning problem where the goal is to categorize new observations into one of the two presented classes. Anomaly detection, on the other hand, tries to solve the problem of identifying rare events or observations in data that don't fit the normal patterns. Both tasks require two distinct classes of samples, and here we use the Clinical Dementia Rating (CDR) scores to divide the samples into ``cognitive normal" or ``AD/Demented."

For binary classification, we will be training various supervised convolutional neural network (CNN)-based models to classify between the two classes. We first perform hyperparameter tuning and record scores for various architectures. We will then use the model with the best performance for domain adaptation. For domain adaptation, we will consider one dataset to be the source distribution and a different dataset to be the target distribution. The model will be trained in an unsupervised domain adaptation setting where the source dataset is labeled and the target data is unlabeled. This will give us good performance metrics for scenarios where we want to use pre-trained models on a set of unlabeled MRI samples. The datasets are publicly available and are sourced from the OASIS database.

Then we look into the problem of anomaly detection, where the samples classified as demented are considered anomalies. before training the models for anomaly detection. We first test various generative autoencoder-based models for image reconstruction and image synthesis. This will give us the performance benchmarks to help pick the best model for both anomaly detection and domain adaptation for anomaly detection. Image reconstruction is an image-to-image operation that deals with the problem of recovering observations from latent space. On the other hand, image synthesis tries to learn the distribution of the data in order to efficiently generate new data samples while maintaining the best distribution and feature variety. We will be studying domain adaptation for anomaly detection in both supervised and unsupervised settings. The unsupervised approach is similar to the previous discussion of binary classification, where the target samples are unlabeled. For the supervised approach, we assume that we have knowledge of the fact that all the available samples are from non-anomalous samples, which is the cognitive normal class. The specific approach used for domain adaptation is ADDA, or adversarial discriminative domain adaptation \cite{tzeng_adversarial_2017}, which will be discussed in detail in the coming sections.

\section{Methodology and Results}
\subsection{Experimental Data}

While there are various publicly available MRI datasets, In particular, OASIS-1 and OASIS-2 datasets from the OASIS database will be used. Neuroimaging datasets are now freely accessible for research thanks to OASIS, also known as the Open Access Series of Imaging Studies. The OASIS-1 dataset is a cross-sectional compilation of 416 participants' T1-weighted MRI scans, ranging in age from 18 to 96. And OASIS-2 is a longitudinal set of 150 individuals' T1-weighted MRI images from 60 to 96 years old. The term "T1-weighted" refers to the timing of radiofrequency pulse sequences used in scanning. Hence, considering the range of ages of the subjects, OASIS-2 is a more balanced dataset when it comes to the distribution of the two classes. We also have access to the demographic data of the subjects for both of the datasets.

\begin{figure}[h] 
\begin{tabular}{ccc}
  \includegraphics[width=1.0in]{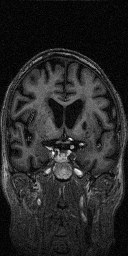} &   \includegraphics[width=2.0in]{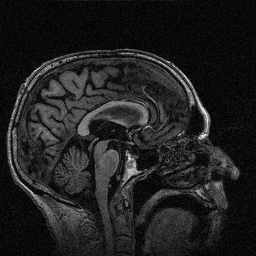} &   \includegraphics[width=1.0in]{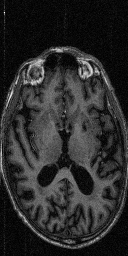}\\
(a) coronal plane & (b) sagittal plane & (c) axial plane\\[6pt]
 \includegraphics[width=1.0in]{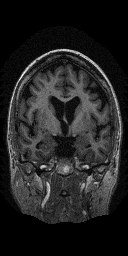} &   \includegraphics[width=2.0in]{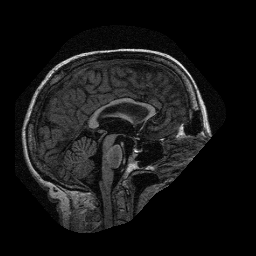} &   \includegraphics[width=1.0in]{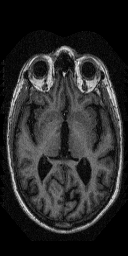}\\
(a) coronal plane & (b) sagittal plane & (c) axial plane\\[6pt]

\end{tabular}
\caption{Data samples of OASIS-2 (top) and OASIS-1 (bottom)}
\label{fig:MRI}
\end{figure}

The demographic data contains information about the Clinical Dementia Rating (CDR) scores. We use the CDR values to first differentiate the samples into "cognitive normal" or "AD/Demented". The CDR is an alternate semi-structured interview that uses scores from six different cognitive and behavioral areas to calculate a number between 0 and 3 for evaluating dementia. We classify samples with a CDR score of 0 as cognitively normal, and samples with a CDR score greater than zero as demented. The samples are normalized based on their mean and standard deviation. The normalized samples are then sliced along the sagittal plane to obtain ten distinct slices of each sample. We use sliced samples due to the computational complexity associated with training the models on whole-brain 3D MRI scans. Figure ~\ref{fig:MRI} shows samples from both the OASIS-1 and the OASIS-2 datasets along the coronal plane, sagittal plane, and axial plane.

\onecolumngrid

\twocolumngrid

\subsection{Binary Classification}

We have explored several models for binary classification in order to get a baseline performance on the source data and compare the models so that we can pick the one with the best performance that can be used for domain adaptation. We used grid search based on the learning rate and the number of epochs. Each data sample contains 10 slices along the sagittal view, so the dimension of each sample is 10x256x256. The models tested are ResNet-18, EfficientNet-B3, ResNeXt-50, and a 3D ResNet-18 model.

The \textbf{ResNet} model \cite{2015arXiv151203385H} was the first one to be tested. To address complexities, deep neural networks are often built with additional layers, which increase accuracy and performance. When more layers are added, the theory behind layering is that they will eventually learn features that are more complex. For example, the first layer might learn to distinguish edges in the case of image analysis, the second layer might learn to recognize textures, the third layer might learn to recognize objects, and so on. Nonetheless, adding too many layers might result in overfitting, vanishing gradients, and other issues. The development of ResNet, or residual networks, which are composed of residual blocks, has solved the complexity associated with training very deep networks. Layers with skipped connections form residual blocks, which are stacked together. The skip connections in ResNet provide a different shortcut way for the gradient to flow through, resolving the issue of vanishing gradients in deep neural networks. The identity functions are learned by the model as a result of these linkages, which helps ensure that the layers that are stacked higher will perform at least as well as the ones present lower. Here, we specifically test a network with 18 layers, which is the ResNet-18 model.

Next, we explored the \textbf{EfficientNet} model \cite{2019arXiv190511946T}. Using a compound coefficient, the convolutional neural network formulation and scaling method EfficientNet uniformly scales all depth, breadth, and resolution dimensions. The EfficientNet scaling method increases network width, depth, and resolution consistently using a set of preset scaling coefficients, in contrast to traditional methods that scale these variables arbitrarily. Network width, depth, and resolution are all uniformly scaled by EfficientNet with the help of a compound coefficient. The rationale behind the compound scaling method is that larger input images require more layers in order to expand the network's receptive field and more channels in order to acquire more fine-grained features on the larger image. We use the EfficientNet-B3 architecture in this instance. In order to determine the relationship between the several fundamental network scaling dimensions while adhering to a set resource limitation, this architecture employs the grid search approach. This could help determine the proper scaling factors for each of the scaled dimensions. The basic network was scaled to the required size based on these computed factors.

Another architecture we tested is the \textbf{ResNeXt-50} model, a ResNeXt model \cite{2016arXiv161105431X} with 50 layers. A ResNeXt replicates a structural component that combines a number of transformations with the same topology. It reveals an additional dimension that is cardinality, or the magnitude of the set of transformations, which is in contrast to a standard ResNet architecture. Thus, the ResNeXt model introduces a new hyperparameter called cardinality to adjust the model's capacity. Additionally, ResNeXt also uses the split-transform-merge paradigm from the Inception model. The ResneXt block's input is projected into a number of lower-dimensional representations instead of applying convolutions over the entire input feature map. We then independently apply only a few convolutional filters to each representation before combining the results. Group convolutions, which were suggested in the AlexNet paper \cite{10.5555/2999134.2999257} as a technique to distribute the convolution computation across numerous GPUs, are pretty similar to this notion. The input is divided channel-wise into groups rather than building filters with the entire channel depth of the input. \textbf{Since the MRI scans are sliced and the slices are passed as channels, each channel of the input is thus uncorrelated with each other, and since ResNeXt divides the input channel-wise into groups, we hypothesize that it could be the ideal architecture for our task.}

Lastly, we evaluate a \textbf{3D-ResNet-18} model. which is essentially a ResNet-18-based model that employs 3D convolution layers. With the earlier architectures, we were passing the slices along the sagittal view as channels, but here we pass the slices as a spatial dimension using 3D convolution to check whether passing the data in this fashion provides any improvement in the performance.

The results of the hyperparameter tuning, i.e., the best model and set of hyperparameters, are presented in the Table ~\ref{table:HTResults} below, and the complete scores are compiled in the appendix. As hypothesized earlier, the ResNeXt model performed the best. Since ResNeXt is the best scoring model, we will be using it as our base architecture for domain adaptation. All models have been built using the PyTorch framework \cite{paszke_pytorch_2019} and trained on an NVIDIA A100 GPU.

\begin{table}[H]
    \centering
    \begin{tabular}{|l|l|l|l|}
    \hline
        Model  & ResNeXt-50 \\ \hline
        Training Epochs  & 50\\ \hline
        Learning Rate  & 2e-4\\ \hline
    \end{tabular}
\caption{Best Modela and Hyperparameters}
\label{table:HTResults}
\end{table}

\begin{table}[H]
    \centering
    \begin{tabular}{|l|l|l|l|}
    \hline
        Model  & ResNeXt-50 \\ \hline
        AUC  & 0.91157\\ \hline
        Accuracy  & 0.84897\\ \hline
        Sensitivity/Recall & 0.88489\\ \hline
        Specificity  & 0.80188\\ \hline
        Precision  & 0.85416\\ \hline
        F1 & 0.86925 \\ \hline
    \end{tabular}
\caption{Supervised Binary Classification on Source (OASIS-2)}
\label{table:BCResults}
\end{table}

\begin{figure}[H]
	\centering
	\includegraphics[width = 3.0in]{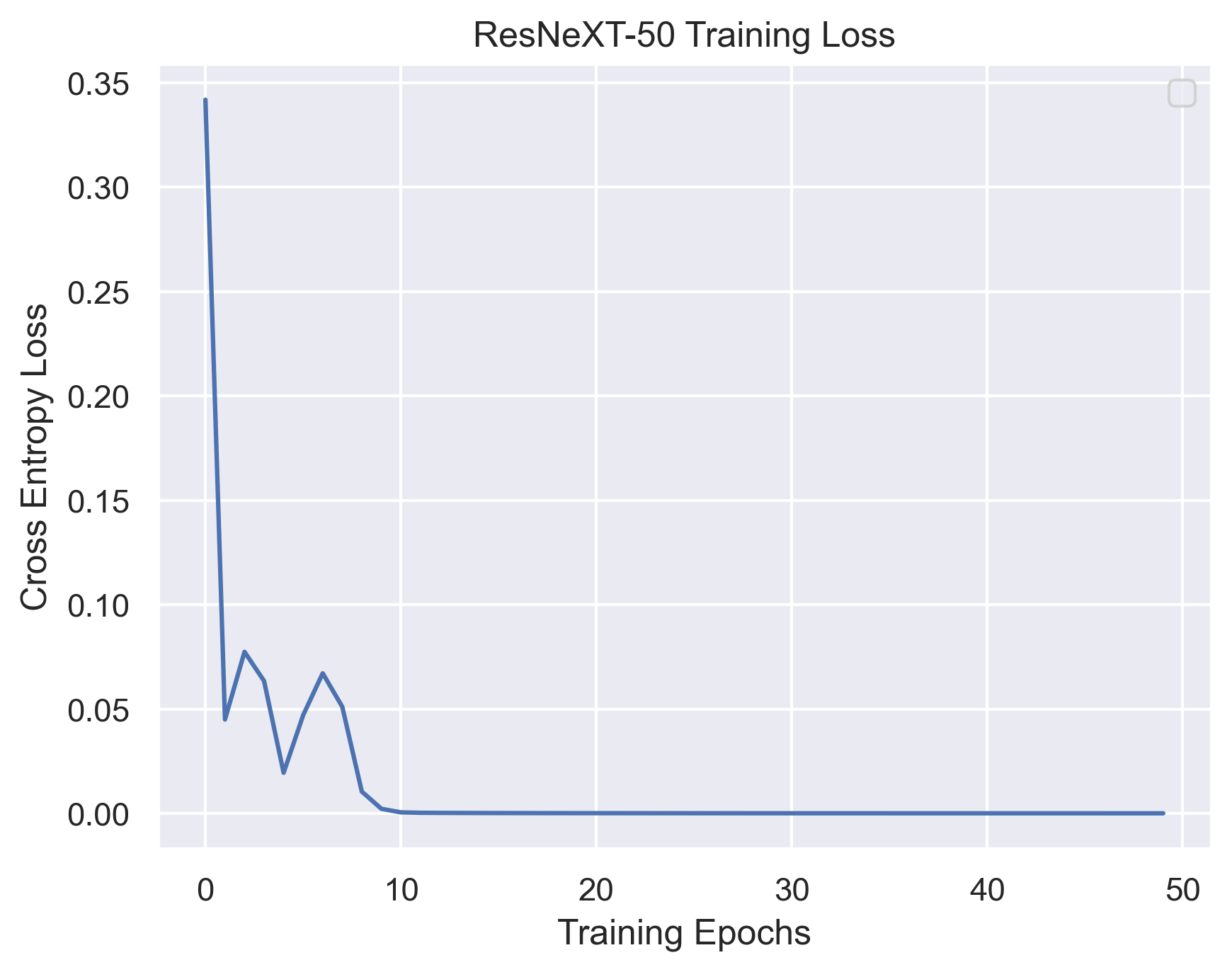}
	\caption{Training Loss of ResNeXt Model}	
 \label{fig:Plot1}
\end{figure}

We will now train the ResNeXt model with this set of hyperparameters on the OASIS-2 dataset, which will be used as our source dataset for domain adaptation. The model is trained with the cross-entropy loss. A classification model's performance is measured by cross-entropy loss, also known as log loss, whose output is a probability value between 0 and 1. For learning, we use a gradient descent optimization algorithm called Adam, or Adaptive Moment Estimation \cite{2014arXiv1412.6980K}. It is an adaptive learning rate method that combines momentum-based stochastic gradient descent with RMSprop. It scales the learning rate using squared gradients, similar to RMSprop, and leverages momentum by using the gradient's moving average rather than the gradient itself, akin to SGD with momentum. The results are presented in the Table ~\ref{table:BCResults}, and Figure ~\ref{fig:Plot1} shows the training loss. The metrics calculated are AUC, accuracy, sensitivity/recall, specificity, precision, and the F1 score.

\subsection{Domain Adaptation}

\begin{figure}[h] 
	\centering
	\includegraphics[width = 6.0in]{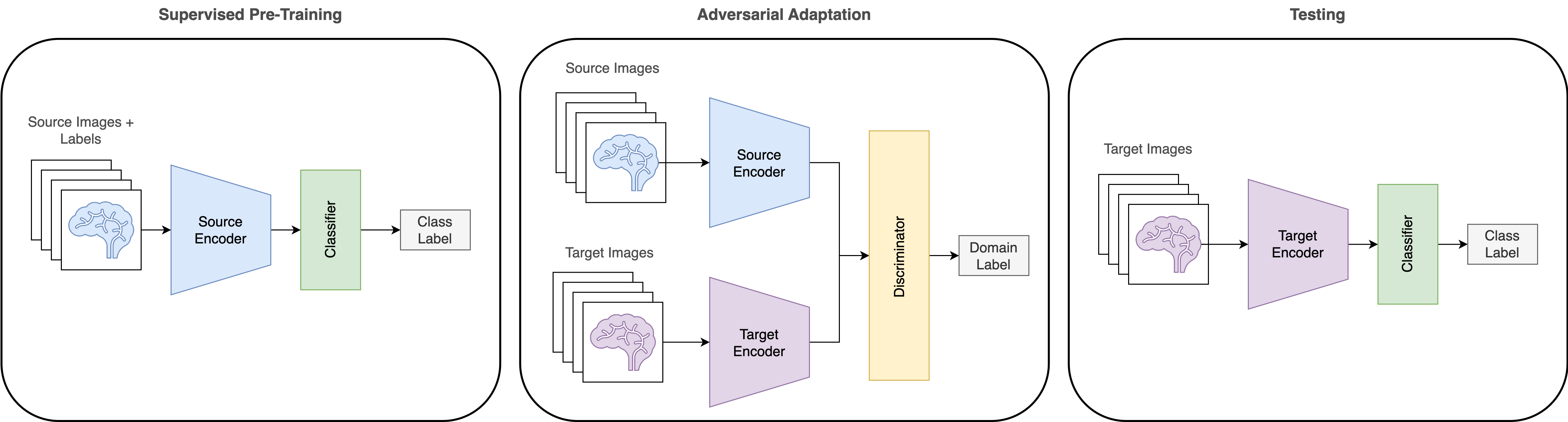}
	\caption{Overview of the ADDA domain adaptation method.}	
  \label{fig:ADDA}
\end{figure}

We will now discuss domain adaptation for binary classification. The ADDA approach is the specific method we employed for domain adaptation. A supervised model will be trained on a source data set with the intention of applying it to a target data set. We employ the ResNeXt-50 as our base architecture for this endeavor, specifically for the encoder, as was previously stated. Referring back to the discussion of the ADDA method, Figure ~\ref{fig:ADDA} gives a description of this method.

ADDA, or Adversarial Discriminative Domain Adaptation, is an adversarial adaptation method with the goal of minimizing the domain discrepancy distance through an adversarial objective with respect to a discriminator. Ideally, the discriminator will be unable to distinguish between the
source and the target distributions. We consider that we have access to source images and labels that come from a source distribution, but when it comes to the target distribution, we have the target images but not the labels. Our objective is to train a target encoder and classifier that can classify the target samples into the source classes. That is done by training the source and target encoders adversarially, and the discriminator tries to distinguish between the mapped latent vectors of the source and target images. The results of domain adaptation are presented in the table ~\ref{table:DAResults}. Surprisingly, the accuracy score obtained for the OASIS-1 dataset via unsupervised domain adaptation is actually higher than some of the state-of-the-art supervised classification results.

\onecolumngrid

\twocolumngrid

\hspace{1cm}

\begin{table}[H]
    \centering
    \begin{tabular}{|l|l|}
    \hline
        Model & ResNeXt-50 \\ \hline
        Accuracy without DA & 0.74624 \\ \hline
        Accuracy ADDA & \textbf{0.8312} \\ \hline
        SOTA Accuracy (Saratxaga et al. \cite{Saratxaga2021-fm}) & \textbf{0.81} \\ \hline
        Sensitivity/Recall & 0.82926 \\ \hline
        Specificity  & 0.83783 \\ \hline
        Precision  & 0.94444 \\ \hline
        F1 & 0.88311 \\ \hline
    \end{tabular}
    \caption{Domain Adaptation, Source: OASIS-2, Target: OASIS-1}
    \label{table:DAResults}
\end{table}

\subsection{Anomaly Detection}

\begin{figure}[H] 
	\centering
	\includegraphics[width = 6.0in]{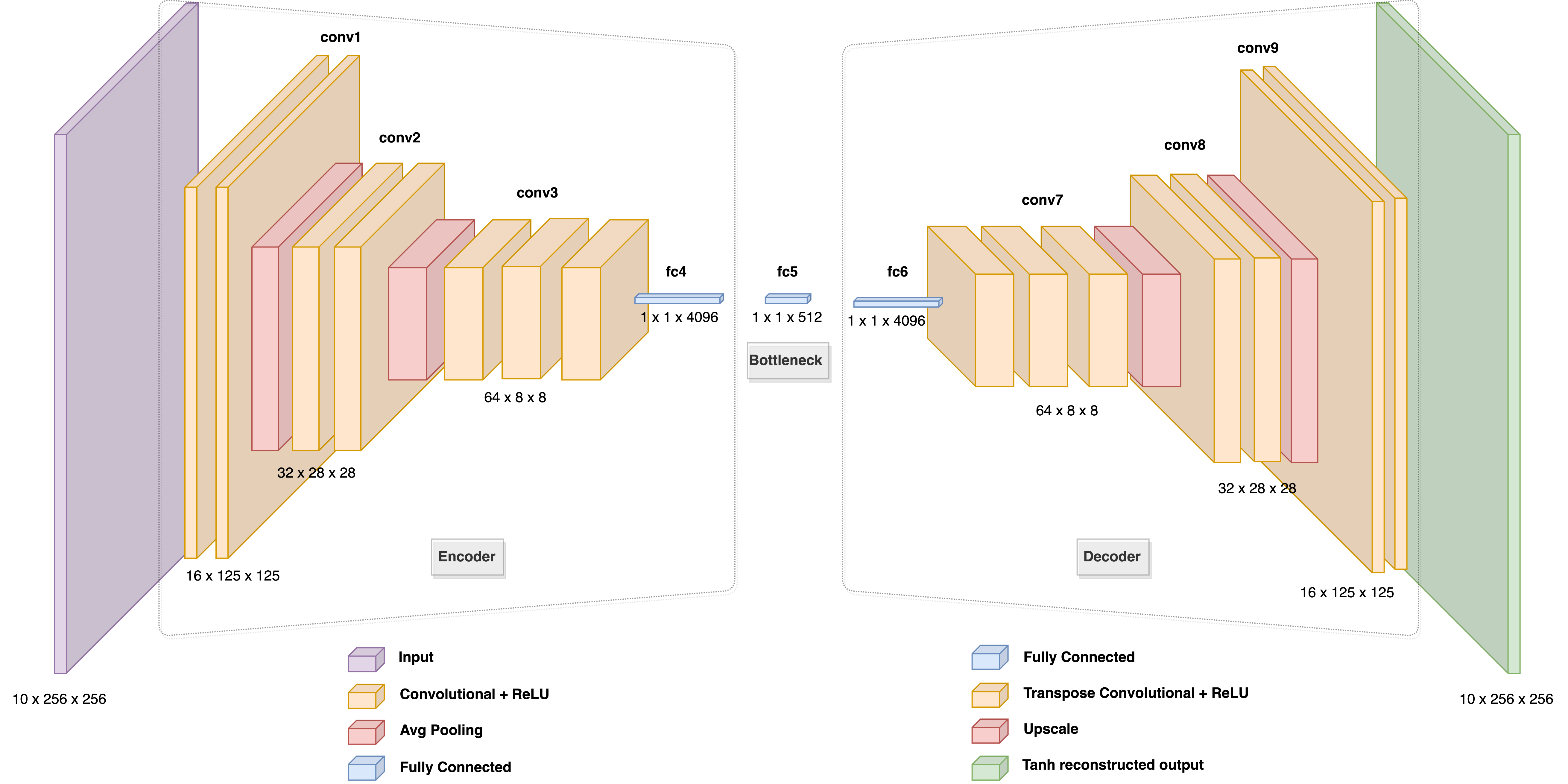}
	\caption{Architecture of the Autoencoder model}		 
   \label{fig:AE}
\end{figure}

In this section, we will be discussing image reconstruction, image synthesis, anomaly detection, and domain adaptation for anomaly detection. We use autoencoder-based generative models such as the adversarial autoencoder \cite{2015arXiv151105644M} and variational autoencoder \cite{2013arXiv1312.6114K} for image reconstruction and image synthesis. This is in order to get a baseline performance of the models so that we can use the best-performing model for anomaly detection and also explore domain adaptation in an anomaly detection setting. The auto-encoder models used consist of an encoder and a decoder network. The decoder network learns to reconstruct the input observations from the latent space after the encoder learns to map the input observations to a latent space with a lower dimension than the input samples. The variational autoencoder imposes an additional constraint on the standard autoencoder in the form of KL divergenceloss , and the adversarial autoencoder replaced this with adversarial learning. The loss function is the mean squared error (MSE) in the case of the adversarial auto-encoder and MSE plus KL divergence for the variational auto-encoder. The optimizer used for training is the Adam optimizer in both cases. The architecture of the autoencoder model has been presented in Figure ~\ref{fig:AE}. We have used Tanh as the output activation.

\onecolumngrid

\twocolumngrid

\begin{figure}[H]
	\centering
	\includegraphics[width = 3.0in]{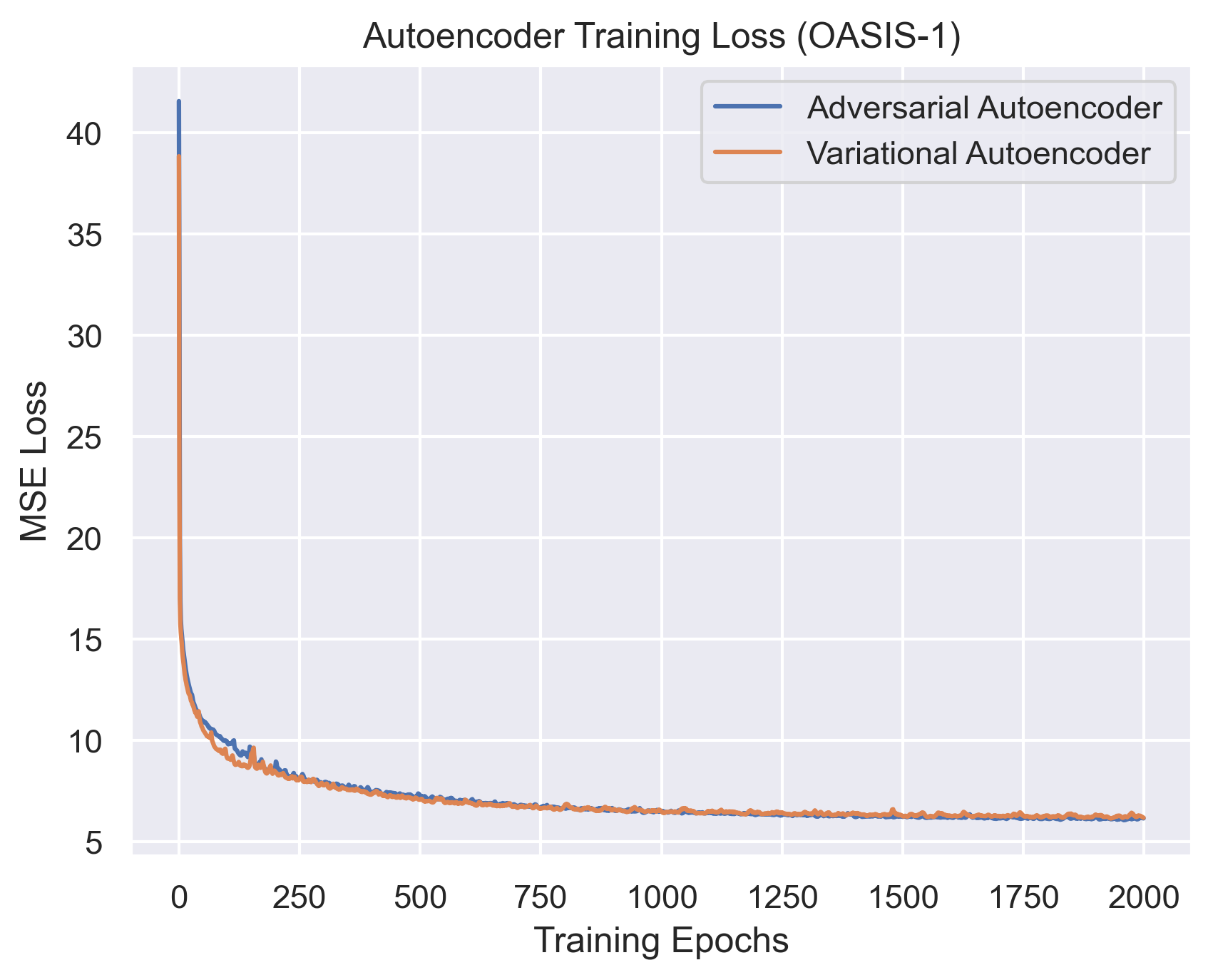}
	\caption{Training Loss of Autoencoder Models on OASIS-1 dataset}	
    \label{fig:plt2}
\end{figure}

\begin{figure}[H]
	\centering
	\includegraphics[width = 3.0in]{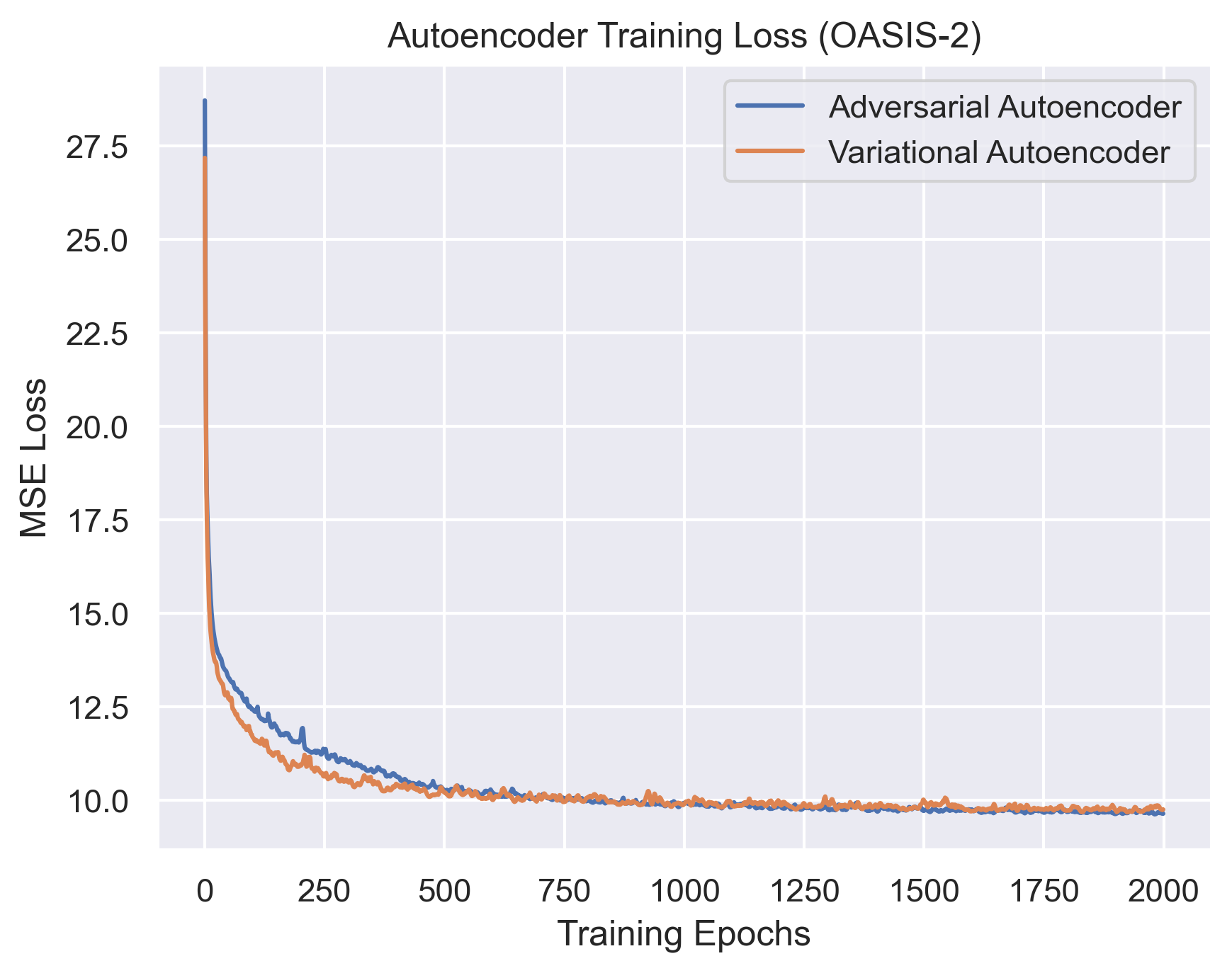}
	\caption{Training Loss of Autoencoder Models on OASIS-2 dataset}	
   \label{fig:plt3}
\end{figure}

Figures ~\ref{fig:plt2}, and ~\ref{fig:plt3} present the training loss of the autoencoder models and the results obtained for image reconstruction and image synthesis are presented in the below Tables ~\ref{table:EMDResults1} and ~\ref{table:EMDResults2}. To test and compare the models quantitatively, we have used the earthmovers' distance (EMD) as the metric. On a high level, EMD measures the distance between two probability distributions. Both models perform similarly, and the OASIS-1 dataset looks to be much more complex than the OASIS-2 dataset.

\onecolumngrid
\hspace{1cm}
\begin{table}[H]
    \centering
    \begin{tabular}{|l|l|l|}
    \hline
        Model & Generated Samples & Reconstructed Samples \\ \hline
        Adversarial Autoencoder & 27.04159&	28.40938 \\ \hline
        Variational Autoencoder & 25.15828	&23.93365 \\ \hline
    \end{tabular}
    \caption{EMD values (OASIS-2)}	
    \label{table:EMDResults1}
\end{table}

\begin{table}[H]
    \centering
    \begin{tabular}{|l|l|l|}
    \hline
        Model & Generated Samples & Reconstructed Samples \\ \hline
        Adversarial Autoencoder & 64.21179 & 63.64729 \\ \hline
        Variational Autoencoder & 66.7877 & 66.27392 \\ \hline
    \end{tabular}
    \caption{EMD values (OASIS-1)}	
    \label{table:EMDResults2}
\end{table}
\twocolumngrid

We now use the trained autoencoder models for anomaly detection. We are considering the demented samples to be anomalies. The basic premise is that the autoencoder models are trained specifically on the cognitively normal samples for image reconstruction, and when the models are tested on the test data samples, they should produce a higher reconstruction loss for the demented samples. We will first test the models on their respective datasets and then use domain adaptation to transfer the model trained on the source dataset, which is OASIS-2, to the target dataset, which is OASIS-1.

The AUC scores calculated based on the reconstruction loss are shown in the Table ~\ref{table:ADResults}, and while the score for the OASIS-1 dataset is poor, we get a decent score for the OASIS-2 dataset, implying that the model can distinguish between cognitively normal and demented samples to some extent. These results should also give us a baseline for domain adaptation.

\begin{table}[H]
\small
\centering
\caption{AUC scores for Anomaly Detection}
\begin{tabular}{| >{\centering\arraybackslash}m{1.2in} | >{\centering\arraybackslash}m{0.8in} |>{\centering\arraybackslash}m{0.8in} |}
\hline
Model & OASIS-1	& OASIS-2
\\
\hline
Adversarial Autoencoder	        &      0.60727&	0.71692\\
Variational Autoencoder	 & 0.67645&	0.68715\\
\hline
\end{tabular}
\label{table:ADResults}
\end{table}

Similar to how we have experimented in the case of binary classification, we will be using the ADDA approach of domain adaptation for anomaly detection. This has been done in two parts: the supervised approach, where we only use the cognitively normal samples of the target dataset, and the unsupervised approach, where we use all the samples of the target dataset from both classes. During our testing, we observed that the adversarial autoencoder model was frequently outperforming the variational autoencoder model, which is why we specifically use the adversarial autoencoder for domain adaptation. 

\onecolumngrid
\hspace{1cm}
\begin{table}[H]
\small
\centering
\caption{AUC scores for Domain Adaptation, Source: OASIS-2, Target: OASIS-1}
\begin{tabular}{| >{\centering\arraybackslash}m{1.2in} | >{\centering\arraybackslash}m{0.8in} |>{\centering\arraybackslash}m{1.8in} |>{\centering\arraybackslash}m{1.8in} |}
\hline
Model	&Without DA&	ADDA (For Supervised Anomaly Detection)	& ADDA (For Unsupervised Anomaly Detection)
\\
\hline
Adversarial Autoencoder	        &      0.78162	&0.81097	&0.73742\\
Variational Autoencoder	 & 0.77341&	0.79150 & 0.71228\\
\hline
\end{tabular}
\label{table:DAADResults}
\end{table}
\twocolumngrid

Examining the results in Table ~\ref{table:DAADResults}, even simply using a naive transfer learning approach that is performing inference using a pre-trained model without domain adaptation improves the AUC score for OASIS-1, but the ADDA approach gives us an improvement in the score, and while using ADDA for anomaly detection in an unsupervised setting resulted in degraded performance, it still outscores the result we got for anomaly detection without any transfer learning or domain adaptation.

\section{Discussion and Future Work}

Although we were able to achieve encouraging results, we have only used the sagittal plane for slicing and training the models, and training the models using samples along other planes could be explored. The anomaly detection results are currently not on par with those obtained for binary classification. However, this approach still has a lot of potential since anomaly detection only requires samples from cognitively normal subjects i.e. we could train the anomaly detection model with a larger set of samples since it’s easier to obtain more samples of cognitively normal subjects as opposed to demented subjects.

Despite the fact that the motivating anomaly detection results show that the auto-encoder models are able to capture the features of the MRI scans, the images generated by them don’t look visually appealing, which is why we chose a more quantitative measure such as EMD to compare the models. Alternative deep generative models, such as generative adversarial networks (GAN), could be explored to improve image synthesis performance.

\section{Conclusion}

In conclusion, we have performed various studies on the OASIS-1 and OASIS-2 datasets. Domain adaptation in the case of binary classification resulted in a significant improvement to the results of the OASIS-1 dataset and outperformed some of the state-of-the-art supervised classification approaches despite the model being trained in an unsupervised setting. In addition, we then explored an alternative way of Alzheimer’s disease detection called anomaly detection and noticed that domain adaptation improved the scores for this application as well. Hence, domain adaptation is a useful technique for studying MRI brain scans and can be used for developing competent methods for Alzheimer’s disease classification.

\section{Acknowledgements}

We would like to thank Dr. Corey Toler-Franklin for her guidance and the University of Florida for providing access to computational resources.

\nocite{*}
\bibliography{Paper}

%\clearpage
%\newpage
%\pagebreak

\onecolumngrid

%\clearpage
%\newpage
\pagebreak

\appendix
\section{Hyperparameter Tuning}

%\clearpage
%\newpage

\begin{table}[H]
\begin{subtable}[c]{1.0\textwidth}
\centering
\begin{tabular}{| >{\centering\arraybackslash}m{1.2in} | >{\centering\arraybackslash}m{1.0in} |>{\centering\arraybackslash}m{1.0in} | >{\centering\arraybackslash}m{1.0in} |>{\centering\arraybackslash}m{1.0in} |}
\hline
LR & Epochs=10 & Epochs=20 & Epochs=50 & Epochs=100
\\
\hline
2.00E-04   & 0.75102 & 0.76745 & 0.78775 & \textbf{0.81632}\\
2.00E-05   & 0.70612 & 0.7102 & 0.72244 & 0.71836\\
\hline
\end{tabular}
\subcaption{ResNet-18}
\end{subtable}

\begin{subtable}[c]{1.0\textwidth}
\centering
\begin{tabular}{| >{\centering\arraybackslash}m{1.2in} | >{\centering\arraybackslash}m{1.0in} |>{\centering\arraybackslash}m{1.0in} | >{\centering\arraybackslash}m{1.0in} |>{\centering\arraybackslash}m{1.0in} |}
\hline
LR & Epochs=10 & Epochs=20 & Epochs=50 & Epochs=100
\\
\hline
2.00E-04   & 0.67755&	0.7551&	0.75918	&\textbf{0.77551}\\
2.00E-05   & 0.70659	&0.70204&	0.72653&	0.7551\\
\hline
\end{tabular}
\subcaption{EfficientNet-B3}
\end{subtable}

\begin{subtable}[c]{1.0\textwidth}
\centering
\begin{tabular}{| >{\centering\arraybackslash}m{1.2in} | >{\centering\arraybackslash}m{1.0in} |>{\centering\arraybackslash}m{1.0in} | >{\centering\arraybackslash}m{1.0in} |>{\centering\arraybackslash}m{1.0in} |}
\hline
LR & Epochs=10 & Epochs=20 & Epochs=50 & Epochs=100
\\
\hline
2.00E-04   & 0.77142&	0.83265	&\textbf{0.85306}&	0.84489\\
2.00E-05   & 0.72653&	0.70612&	0.7102&	0.70612\\
\hline
\end{tabular}
\subcaption{ResNeXt-50}
\end{subtable}

\begin{subtable}[c]{1.0\textwidth}
\centering
\begin{tabular}{| >{\centering\arraybackslash}m{1.2in} | >{\centering\arraybackslash}m{1.0in} |>{\centering\arraybackslash}m{1.0in} | >{\centering\arraybackslash}m{1.0in} |>{\centering\arraybackslash}m{1.0in} |}
\hline
LR & Epochs=10 & Epochs=20 & Epochs=50 & Epochs=100
\\
\hline
2.00E-04   & 0.6122&	0.53061	&0.66938&	0.59591\\
2.00E-05   & 0.70612	&\textbf{0.70804}	&0.68979	&0.6853\\
\hline
\end{tabular}
\subcaption{3D ResNet-18}
\end{subtable}

\caption{Hyperparameter Tuning, Accuracy scores for Binary Classification (OASIS-2)}
\end{table}

\begin{table}[H]
\begin{subtable}[c]{1.0\textwidth}
\centering
\begin{tabular}{| >{\centering\arraybackslash}m{1.2in} | >{\centering\arraybackslash}m{1.0in} |>{\centering\arraybackslash}m{1.0in} | >{\centering\arraybackslash}m{1.0in} |>{\centering\arraybackslash}m{1.0in} |}
\hline
Epochs=10 & Epochs=20 & Epochs=30 & Epochs=50 & Epochs=100
\\
\hline
0.75	&\textbf{0.80125}	&0.78874	&0.775	&0.76625\\
\hline
\end{tabular}
\subcaption{ResNeXt-50, Critic LR: 1e-5, Target LR: 1e-6}
\end{subtable}

\caption{Hyperparameter Tuning, Accuracy scores for Domain Adaptation (Source: OASIS-2, Target: OASIS-1)}
\end{table}

%\pagebreak
%\clearpage
\end{document}